\documentclass{ws-ijmpa}

\begin{document}

\markboth{A.~A.Banishev  {\it et al.}}
{Observation of Reduction in Casimir Force}

%
\catchline{}{}{}{}{}
%

\title{OBSERVATION OF REDUCTION IN CASIMIR FORCE WITHOUT CHANGE
OF DIELECTRIC PERMITTIVITY}

\author{ A.~A.~BANISHEV,
C.-C.~CHANG, R.~CASTILLO-GARZA,
}

\address{Department of Physics and
Astronomy, University of California, Riverside, CA 92521,
USA}

\author{G.~L.~KLIMCHITSKAYA}

\address{North-West Technical University, Millionnaya Street 5,
St.Petersburg, 191065, Russia}

\author{V.~M.~MOSTEPANENKO}

\address{Noncommercial Partnership ``Scientific Instruments'',
Tverskaya Street 11, Moscow, 103905, Russia}

\author{U.~MOHIDEEN}

\address{Department of Physics and
Astronomy, University of California, Riverside, CA 92521,
USA {\protect \\} Umar.Mohideen@ucr.edu}
\maketitle

\begin{history}
\received{17 November 2011}
\revised{4 January 2012}
\end{history}

\begin{abstract}
Additional information is provided on the effect of the significant
(up to 35\%) reduction in the magnitude of  the
Casimir force between an Au-coated sphere and an
indium tin oxide film which was observed after UV treatment of the
latter. A striking feature of this effect is that the reduction
is not accompanied with a corresponding
 variation of the dielectric permittivity,
as confirmed by direct ellipsometry measurements.
The measurement data are compared with computations using the Lifshitz
theory. It is shown that the data for the untreated sample are in a very
good agreement with theory taking
into account the free charge carriers in the indium
tin oxide. The data for the UV-treated sample exclude the theoretical
results obtained with account of free charge carriers. These data are found
to be in a very good agreement with theory disregarding the free charge
carriers in an indium tin oxide film.
A possible theoretical explanation of our observations as a result of
phase transition of indium tin oxide from metallic to dielectric
state is discussed in comparison with other related experiments.

\end{abstract}
\ccode{PACS numbers: 78.20.-e, 78.66.-w, 12.20.Fv, 12.20.Ds}

\section{Introduction}

In the last few years experiments on the Casimir effect have been
numerous and varied. They deal with different materials, such as
metals, semiconductors and dielectrics, and with the influence of
phase transitions on the Casimir force.\cite{1}\cdash\cite{4}
Many experiments tried to significantly reduce the magnitude of
the Casimir force, as compared\cite{5}\cdash\cite{7} to
two test bodies made of good metal, such as Au.
This was achieved by measuring the Casimir force between an Au
sphere and a plate made of different semiconductors,\cite{8}\cdash\cite{12}
semimetals,\cite{13} and when the plate materials undergo
transformation from metallic to dielectric state\cite{14,15} or from
crystalline to amorphous phase.\cite{16}
In all the cases considered so far any detectable changes in the magnitude
of the Casimir force were accompanied by significant variations in the
dielectric permittivity of the plate over a wide region of imaginary
frequencies. This observation is commonly considered as a natural and
invariable consequence of the Lifshitz theory.\cite{1,17}

Here, we report additional information on a significant (up to 35\%)
reduction in the magnitude of  the Casimir force between an Au-coated
sphere and an indium tin oxide (ITO) film recently observed\cite{18}
which occurs  after UV treatment of the ITO. A striking feature of
this phenomenon is that the reduction in the force magnitude is not
accompanied with corresponding variations in the dielectric permittivity
of  ITO under UV treatment. This is confirmed by the direct
ellipsometry measurements of the imaginary parts of the dielectric
permittivity for both the untreated and UV-treated samples.

ITO (In${}_2$O${}_3$:Sn)
has long been suggested\cite{19} for use in Casimir physics
due to its unique electric properties. It was measured\cite{11,12}
that the gradient of the Casimir force between an Au sphere and an
ITO plate is roughly 40\%--50\% smaller than between an Au sphere and
an Au plate. The same rate of a decrease, as compared to the case of
Au-Au test bodies, is confirmed by us\cite{18} for the magnitude
of the Casimir force between an Au sphere and an untreated ITO sample.
The overall decrease for a UV-treated sample is up to 65\% in
comparison to Au-Au surfaces.

All figures presented in this paper use the experimental data of the
second set of our measurements performed before and after the
UV treatment (in previously published paper\cite{18} only the first
measurement set has been used). Here we report for the first time the
experimental data for the total measured force (electric plus
Casimir) and perform detailed comparison between experiment and
theory over a narrower separation region from 60 to 150\,nm
where the measurement data are the most precise.
The measurement data for the untreated sample are found to be in a very
good agreement with the Lifshitz theory when the contribution of free
charge carriers in the ITO is taken into account. For the UV-treated
sample, the measurement data agree very well with computations
disregarding free charge carriers in the ITO film. These data are in
complete disagreement with theory taking
into account the contribution of free
charge carriers in the dielectric permittivity of ITO.
We discuss in
detail a possible theoretical explanation for the observed new effect
of reduction in the Casimir force without change of dielectric
permittivity. We explain this effect by the phase transition from
metallic to dielectric state which occurs in the ITO film under the
influence of UV treatment. We compare our measurements with the
results of other experiments (already performed and proposed)
where the magnitude of the Casimir force was changed as a result
of the phase transition.

The paper is organized as follows. In Sec.~2 we present the
experimental results for the Casimir force, the dielectric
permittivity of ITO and the surface roughness.
Section~3 contains comparison between experiment and
theory. Theoretical explanation of the effect of reduction in
Casimir force without change in dielectric permittivity is considered
in Sec.~4. Section~5 contains our conclusions and discussion.

\section{Experimental Results}

We used a modified multimode atomic force microscope (AFM) to measure
the total force (electrostatic plus Casimir) between an Au-coated
polystyrene sphere and an ITO film
deposited on a quartz plate. Measurements were performed at a pressure
of $10^{-6}\,$Torr at $2^{\circ}$C.
The radius of the Au-coated sphere was measured to be
$R=101.2\pm 0.5\,\mu$m. The Au coating was done in an oil free thermal
evaporator at a very low deposition rate of 3.37{\AA}/min.
The thickness of the Au layer was $105\pm 1\,$nm.
The ITO film was prepared by RF sputtering (Thinfilm Inc.).
The thickness and nominal resistivity of the ITO film were measured to
be $74.6\pm 0.2\,$nm and $42\,\Omega$/sq, respectively.
The ITO sample was cleaned in an ultrasonic bath
with acetone then with methanol and finally with ethanol alternated
with rinsing in DI water, and finally dried in a flow of nitrogen gas.

After the force measurements for the untreated ITO sample were completed,
it was UV-treated for 12 hours in a special chamber containing a UV
lamp. The spectrum of this lamp has the primary peak at 254\,nm
($5.4\,\mbox{mW/cm}^2$ at 1.9\,cm distance) and a secondary peak at 365\,nm
($0.2\,\mbox{mW/cm}^2$ at the same distance).
After the UV treatment the sample was cleaned as described above.
To stabilize the laser used for the detection of deflection of the AFM
cantilever, we employed a liquid nitrogen cooling system, which
maintained the temperature at $2^{\circ}$C.

\subsection{Measurement of the Casimir force between an Au sphere and
an ITO plate}

The scheme of force measurements using an AFM is as
follows.\cite{1,2,4,5,8}\cdash\cite{16}
An Au-coated cantilever of an AFM with attached sphere undergoes
a deflection $z$ in response to the sphere-plate force $F_{\rm tot}$
in accordance with Hooke's law $F_{\rm tot}=kz$, where $k$ is the spring
constant. The photodetectors measure this deflection through
differential measurement of the laser beam intensity reflected from the
cantilever. The respective deflection signal $S_{\rm def}$ as a function
of separation distance $a$ between the sphere and the plate leads to
a force-distance curve. This signal can be calculated according to
$z=mS_{\rm def}$ and measured in volts, where $m$
measured in nm/V is the cantilever
deflection per unit photodetector signal, also called the deflection
coefficient.

Due to the presence of surface roughness, there is some nonzero
minimum separation $z_0$ that can be achieved when the sphere and the
plate are approaching. It is called the separation on contact.
Taking into account deflection of the cantilever under the influence of the
total force, the absolute separation between the sphere and the plate is
given by
\begin{equation}
a=z_0+z_{\rm piezo}+mS_{\rm def}.
\label{eq1}
\end{equation}
\noindent
Here, $z_{\rm piezo}$ is the distance traveled by the plate attached
to the top of the piezoelectric actuator which was calibrated
interferometrically.\cite{20,21} To change the separation, a continuous
triangular voltage was applied to the actuator. As a result, an ITO
plate was moved towards the Au sphere starting at maximum separation
of $2\,\mu$m until the contact, and the cantilever deflection was
recorded at every 0.2\,nm.

The experimental Casimir force was found from the equation
\begin{equation}
F(a)=F_{\rm tot}(a,V_i)-F_{\rm el}(a,V_i)=
\tilde{k}S_{\rm def}(a,V_i)-F_{\rm el}(a,V_i),
\label{eq2}
\end{equation}
\noindent
where $V_i$ ($i=1,\,2,\,\ldots,\,10$) are voltages applied to the ITO
plate while the sphere remained grounded, and $\tilde{k}=km$.
The voltage supply used had a $1\,\mu$V resolution, and special
measures were undertaken to prevent surge currents and to reduce
electrical noise. The electric force between a conducting plate and
a conducting sphere takes the form\cite{1,22}
\begin{equation}
F_{\rm el}(a,V_i)=X(a)(V_i-V_0)^2,
\label{eq3}
\end{equation}
\noindent
where $V_0$ is the residual potential difference due to different
work functions of the sphere and plate materials.
$X(a)$ is the known function of $a$ and $R$. In the wide range of
separations it can be presented with less than 0.01\% error in the
polynomial form\cite{9}
\begin{equation}
X(a)=-2\pi\epsilon_0\sum_{i=-1}^{6}c_i\left(
\frac{a}{R}\right)^i,
\label{eq4}
\end{equation}
\noindent
where $c_i$ are numerical coefficients.\cite{9}

Equations (\ref{eq1})--(\ref{eq4}) can be used to perform the electrostatic
calibrations, i.e., to determine the values of $V_0$, $m$, $z_0$ and
$\tilde{k}$ by applying voltages $V_i$ to the ITO film and by
measuring the respective deflection signal $S_{\rm def}$.
The calibration process is discussed in detail in Refs.~\refcite{18,23}
and \refcite{24}. {}From the tops of the parabolic dependences of
$S_{\rm def}$ on $V_i$ the mean values of
$V_0=-196.8\pm 1.5\,$mV and $V_0=64.8\pm 2.0\,$mV for the untreated and
UV-treated samples, respectively, were determined in the second
measurement set considered here.
To find the mean values of $V_0$, we first determined the individual $V_0$
and found that they are separation-independent in the limits of random
errors if the corrections due to mechanical drift of the sphere-plate
separation and to finiteness of the data acquisition rate are introduced.
Thus, our measurements are free of anomalies discussed in the
literature\cite{25}\cdash\cite{28} and do not support the existence
in our setup of
large electric fields due to patch effect,\cite{29} surface
contaminants etc. in addition to the electrostatic force (\ref{eq3}).
Next, from the curvatures of parabolas the mean values of
other parameters were determined related to the second measurement set:
$m=104.4\pm 0.5\,$nm/V,
$z_0=29.6\pm 0.5\,$nm,
$\tilde{k}=1.51\pm 0.02\,$nN/V
for the untreated sample and
$m=104.2\pm 0.6\,$nm/V,
$z_0=29.0\pm 0.6\,$nm,
$\tilde{k}=1.51\pm 0.02\,$nN/V
for the UV-treated sample.
The error bars in the values of all the above parameters are
indicated at a 95\% confidence level.

\begin{figure*}[t]
\vspace*{-7.7cm}
\centerline{\hspace*{3.2cm}\psfig{file=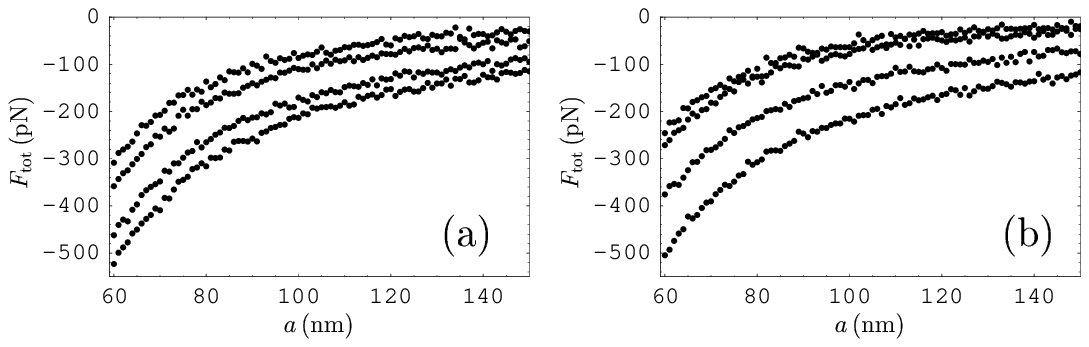,width=24cm}}
\vspace*{-22.5cm}
\caption{Total measured forces as functions of separation with
different applied voltages are indicated as dots for
(a) the untreated and (b) UV-treated sample (see text for further
discussion).}
\end{figure*}

The total force $F_{\rm tot}$ as a function of separation was measured
10 times with a step of 0.2\,nm for each of the 10 different voltages
applied to the plate.
Note that only the interpolated data at 1\,nm step were analyzed.
As an example in Fig.~1  we demonstrate four
typical force-distance curves from bottom to top (a) for the untreated
sample with applied voltages --265\,mV, --255\,mV, --230\,mV, --200\,mV,
respectively, and (b) for the UV-treated sample with respective
applied voltages equal to 140\,mV, 120\,mV, 90\,mV, and 70\,mV.
As can be seen in Fig.~1(a,b), for the
 largest magnitude of applied voltages
(bottom curves) the total measured forces are rather close in
magnitude. For the applied voltages --265\,mV and 140\,mV the
magnitudes of $V-V_0$ are equal to approximately 68 and 75\,mV
for the untreated and UV-treated samples, respectively. These give
major contributions to the magnitudes of the total force.
Thus, without detailed analysis (see Sec.~3) it is not possible to
make any conclusion about the magnitudes of the Casimir force.
A different situation arises with the top curves in Fig.~1(a,b)
[the applied voltages are equal to (a) --200\,mV and (b) 70\,mV].
Here $|V-V_0|$ is equal to only 3 and 5\,mV for the untreated and
UV-treated samples, respectively. The magnitudes of the respective
electric forces are rather small in comparison with still large
(and quite different) magnitudes of the total forces shown by the
top curves in Fig.~1(a,b). This suggests that the Casimir forces
from the untreated and UV-treated samples giving major contributions
to the total forces shown by the top curves are significantly different
(a conclusion confirmed in Sec.~3).

\begin{figure*}[t]
\vspace*{-7.7cm}
\centerline{\hspace*{3.2cm}\psfig{file=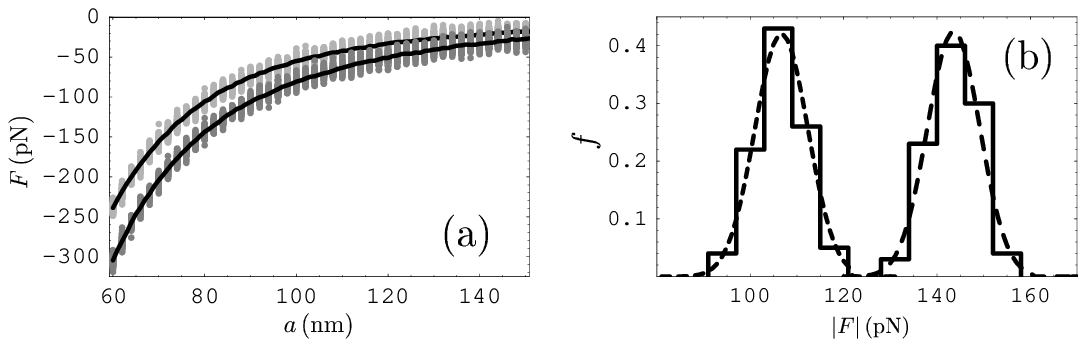,width=24cm}}
\vspace*{-22.5cm}
\caption{(a) Mean measured Casimir forces as a function of separation
are shown by the lower and upper solid lines  for
the untreated and UV-treated sample, respectively.
Respective 100 individual values of the force are shown as dark-grey
and light-grey dots at 2\,nm intervals.
(b) The histograms for measured Casimir force for
the untreated (right) and UV-treated (left) sample at $a=80\,$nm.
$f$ is the fraction of 100 data points having the force values in
the bin indicated by the vertical lines.}
\end{figure*}
Using Eqs.~(\ref{eq2})--(\ref{eq4}), at each separation 100 values of
the Casimir force were obtained shown in Fig.~2(a) by the dark-grey and
light-grey dots with a step of 2\,nm for the untreated and UV-treated sample,
respectively. The respective black
solid lines show the mean Casimir forces
averaged over 100 repetitions. From Fig.~2(a) it is seen that the
UV treatment results in a significant decrease in the magnitude of the
Casimir force from 21\% to 35\% depending on separation. In Fig.~2(b)
we present a histogram for the measured Casimir force at $a=80\,$nm for
the untreated sample (right) and UV-treated sample (left).
The corresponding Gaussian distributions are shown by the dashed lines.
They are characterized by the mean Casimir forces equal to
$F=-143.7\,$pN and $F=-105.5\,$pN for the untreated and UV-treated
{}From Fig.~2(b) it is seen that the Gaussian distributions for the
untreated and UV-treated samples do not overlap lending great confidence
to the effect of reduction of the Casimir force.
{}From the lower solid line in Fig.~2(a) one can conclude also that the
use of an untreated ITO plate leads to a 40\%--50\% decrease in the
force magnitude in comparison with the case of two Au bodies
in agreement with previous work\cite{11,12} where a similar result
was obtained for the Casimir pressure.

\begin{figure*}[b]
\vspace*{-7.7cm}
\centerline{\hspace*{3.2cm}\psfig{file=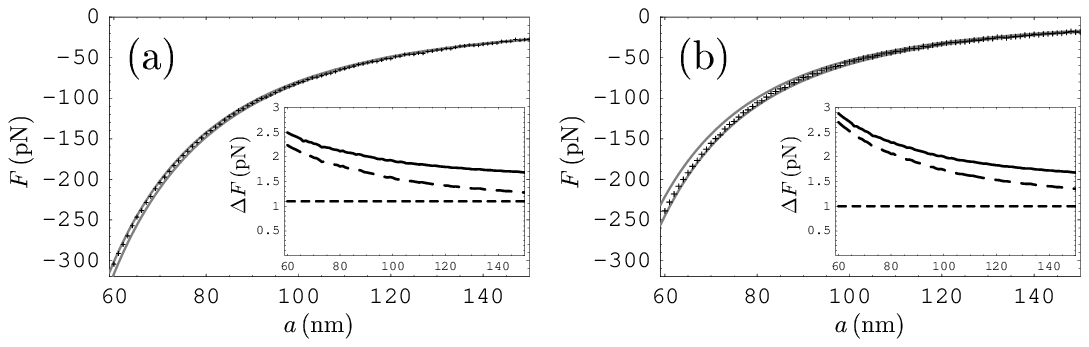,width=24cm}}
\vspace*{-22.5cm}
\caption{The mean measured Casimir force indicated as crosses and
the theoretical Casimir force shown by the pairs of solid lines
as functions of separation  for
(a) the untreated (free charge carriers are included) and
(b) UV-treated sample (free charge carriers are omitted).
In the insets the short-dashed, long-dashed and solid lines show
the random, systematic and total experimental errors, respectively,
for the same samples as in the main fields.}
\end{figure*}
In Fig.~3(a,b) the mean measured Casimir forces are shown as crosses
for the untreated and UV-treated samples, respectively.
The arms of the crosses indicate the total experimental errors in
$a$ and $F$ calculated at a 95\% confidence level. The error in $a$
is mostly determined by the error in $z_0$
indicated above. The total experimental
errors in the Casimir force, $\Delta F$, as a function of $a$,
are shown by the solid lines in the insets to Fig.~3(a,b).
They are obtained by adding in quadrature the random and systematic
erros shown in the insets by the short-dashed and long-dashed lines,
respectively. As can be seen in the insets, the random errors do not
depend on separation. The systematic errors shown in the insets to
Fig.~3(a,b) were found as a combination of the systematic error in the
total measured force and the error in the electrostatic force
subtracted in accordance with
Eq.~(\ref{eq2}). In its turn, the systematic error in
the total measured force is determined by the instrumental noise including
the background noise level, and the errors in calibration.
The error in the calculation of the electric force plays the role of
a systematic error with respect to the Casimir force obtained from
Eq.~(\ref{eq2}). It is mostly determined by the error in the
measurement of separations. In fact the error in the electric force
depends on the applied voltage (to obtain the long-dashed lines in
the insets, the mean values of this error averaged over
10 applied voltages were used).

\subsection{Investigation of the dielectric permittivity
 of ITO and surface  roughness}

\begin{figure*}[b]
\vspace*{-7.7cm}
\centerline{\hspace*{3.2cm}\psfig{file=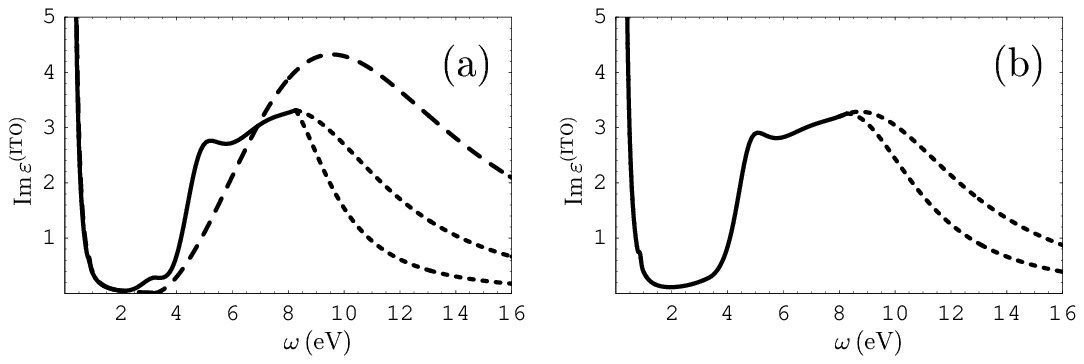,width=24cm}}
\vspace*{-22.5cm}
\caption{The imaginary parts of dielectric permittivity of ITO
measured by ellipsometry as functions of frequency are shown by
the solid lines for (a) the untreated and (b) UV-treated sample.
The short-dashed lines present possible extrapolations of the
measured data to higher frequencies. The long-dashed line presents
the results of Ref.~30.}
\end{figure*}
The optical properties of ITO depend on a layer composition, thickness
and other parameters. Previous experiments\cite{11,12} used the
parametrization of the dielectric permittivity of ITO obtained\cite{30}
with the help of the Tauc-Lorentz model\cite{31} and the Drude model.
This did not lead to good agreement between experiment and theory.
For this experiment the imaginary parts of the dielectric permittivity
of ITO were determined by means of ellipsometry\cite{32} for both
untreated and UV-treated samples. The region of frequencis from 0.04 to
8.27\,eV was covered by the IR-VASE and VUV-VASE ellipsometers.
The obtained results are shown in Fig.~4(a,b) by the solid lines.
(Note that the ITO resistivity decreases with depth leading to the
so-called {\it top} and {\it bottom} dielectric permittivities differing
in the region $\omega<0.5\,$eV.)
In the region of frequencies $\omega<0.04\,$eV the measured top
${\rm Im}\,\varepsilon^{\rm (ITO)}$
which was found to lead to a good agreement with measured Casimir forces
was extrapolated using the imaginary part
of the Drude model with the plasma frequency $\omega_p=1.5\,$eV and
the relaxation parameter $\gamma=0.128\,$eV (for the untreated sample)
and $\omega_p=1.5\,$eV, $\gamma=0.132\,$eV (for the UV-treated sample).
For comparison purposes
the dielectric permittivity\cite{30} is shown in Fig.~4(a) by the
long-dashed line.

The extrapolation of the measured data to higher frequencies
(up to 16\,eV) is also required keeping in mind that the Casimir force
was measured at rather short separations $a\geq 60\,$nm.
This was done by means of the oscillator function
\begin{equation}
{\rm Im}\,\varepsilon^{\rm (ITO)}(\omega)=
\frac{g_0\gamma_0\omega}{(\omega^2-\omega_0^2)^2+\gamma_0^2\omega^2}.
\label{eq5}
\end{equation}
\noindent
For the untreated sample, the reasonable smooth extrapolations are
bounded between the short-dashed lines in Fig.~4(a) which are
described by Eq.~(\ref{eq5}) with the parameters
$g_0=240.54\,\mbox{eV}^2$, $\gamma_0=8.5\,$eV, $\omega_0=9.0\,$eV
(the upper line) and
$g_0=111.52\,\mbox{eV}^2$, $\gamma_0=4.0\,$eV, $\omega_0=8.0\,$eV
(the lower line).
For the UV-treated sample [Fig.~4(b)] the corresponding parameters are
$g_0=280.28\,\mbox{eV}^2$, $\gamma_0=9.2\,$eV, $\omega_0=9.8\,$eV
(the upper short-dashed line) and
$g_0=128.28\,\mbox{eV}^2$, $\gamma_0=4.5\,$eV, $\omega_0=8.8\,$eV
(the lower short-dashed line).
{}From Fig.~4(a,b) it is seen that the UV treatment does not lead to
any significant changes in the
${\rm Im}\,\varepsilon^{\rm (ITO)}(\omega)$.
Using the Kramers-Kronig relation, we have obtained
$\varepsilon^{\rm (ITO)}(i\xi)$ for both the untreated and UV-treated
samples. The results are shown by the pairs of solid and dashed lines
in Figs.~5(a) and 5(b) for the untreated and UV-treated samples,
respectively. Each pair of lines corresponds to the pair of
short-dashed lines in Fig.~4(a,b). The dashed lines in Fig.~5(a)
[solid lines in Fig.~5(b)] present $\varepsilon^{\rm (ITO)}(i\xi)$
with the contribution of free charge carriers omitted.
{}From Fig.~5(a,b) it is also seen that $\varepsilon^{\rm (ITO)}(i\xi)$
is scarcely affected by the UV treatment.
\begin{figure*}[t]
\vspace*{-7.7cm}
\centerline{\hspace*{3.2cm}\psfig{file=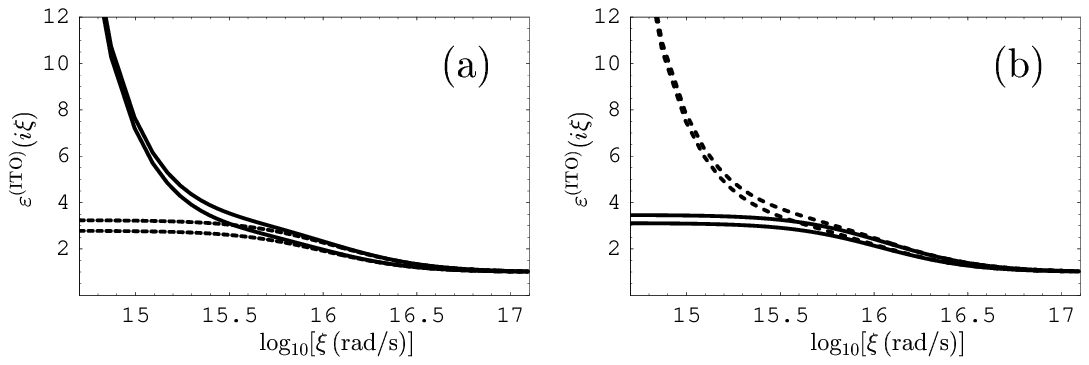,width=24cm}}
\vspace*{-22.5cm}
\caption{The dielectric permittivity of ITO
as functions of imaginary frequency  for (a) the untreated and
(b) UV-treated sample.
The two solid and two dashed lines are obtained with different
extrapolations of the ellipsometry data to higher frequencies
shown by the short-dashed lines in Fig.~4(a,b).
The solid and dashed lines correspond to (a) included and
omitted contribution of free charge carriers, respectively,
and (b) omitted and included
contribution of free charge carriers, respectively.}
\end{figure*}

The averaged dielectric permittivity of the quartz plate underlying
the ITO film was described in the Ninham-Parsegian approximation
\begin{equation}
\varepsilon^{(Q)}(i\xi)=1+\frac{C_{\rm IR}}{1+
\frac{\xi^2}{\omega_{\rm IR}^2}}+\frac{C_{\rm UV}}{1+
\frac{\xi^2}{\omega_{\rm UV}^2}}
\label{eq6}
\end{equation}
\noindent
with the parameters\cite{33}
$C_{\rm IR}=1.93$, $C_{\rm UV}=1.359$,
$\omega_{\rm IR}=0.1378\,$eV, and $\omega_{\rm UV}=13.38\,$eV.
As to $\varepsilon^{\rm (Au)}(i\xi)$, it was obtained by means of
the Kramers-Kronig relation from the tabulated optical data\cite{34}
extrapolated to low frequencies using the imaginary part of the
Drude dielectric permittivity with $\omega_p=9.0\,$eV and
$\gamma=0.035\,$eV. It was recently shown\cite{35} that such
extrapolation nicely fits the optical data measured over a wide region
of frequencies.

\begin{figure*}[t]
\vspace*{-5.5cm}
\centerline{\hspace*{2.6cm}\psfig{file=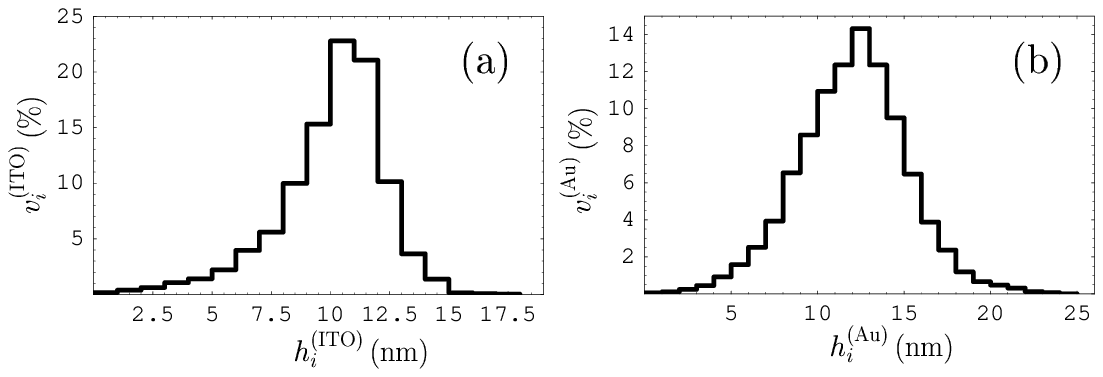,width=23.7cm}}
\vspace*{-24.cm}
\caption{The fractions of the area $v_i$ covered with
roughness of heights $h_i$ for (a) ITO and
(b) Au surfaces.
}
\end{figure*}
Using an AFM we have measured the roughness profiles on both the
ITO sample and the Au-coated sphere. Specifically, the fractions of
area $v_i^{\rm (ITO)}$ with heights $h_i^{\rm (ITO)}$
(where $i=1,\,2,\,\ldots,\,18$) on an ITO film and
$v_i^{\rm (Au)}$ with heights $h_i^{\rm (Au)}$ on an Au sphere
(where $i=1,\,2,\,\ldots,\,25$)
were determined [see Figs.~6(a) and 6(b), respectively].
These heights are measured from the absolute minimum level on the
respective test body $h_1^{\rm (ITO)}=h_1^{\rm (Au)}=0$.
The zero roughness levels on ITO and Au surfaces, relative to which
the mean values of the roughness are equal to zero,\cite{1,2}
take the values $H_0^{\rm (ITO)}=9.54\,$nm and
$H_0^{\rm (Au)}=11.51\,$nm.
This information allows computations of the Casimir force with account
of surface roughness.

\section{Comparison Between Experiment and Theory}

The Casimir force $F$ acting between a smooth Au-coated sphere and
a smooth ITO film deposited on a quartz plate was calculated by using
the Lifshitz theory\cite{17} adapted\cite{1} for a four-layer planar
system (Au-vacuum-ITO-quartz) and the proximity force approximation.
The latter is valid\cite{36} for the experimental parameters
and leads to only a very small error of about $a/R\sim 0.1$\%.
The thicknesses of the Au layer and the quartz plate are quite
sufficient to model them as semispaces. In this way the force $F$ was
found as a function of separation using the dielectric permittivities
specified in Sec.~2.2. Computations were performed at the laboratory
temperature equal to $2^{\circ}$C.
To obtain the theoretical Casimir forces $F^{\rm theor}$ acting
between rough surfaces we have averaged $F$ over the roughness profiles
shown in Fig.~6(a,b)
using the method of geometrical averaging:\cite{1,2}
\begin{equation}
F^{\rm theor}(a,T)=\sum_{i=1}^{18}\sum_{k=1}^{25}
v_i^{\rm (ITO)}v_k^{\rm (Au)}F(a+H_0^{\rm (ITO)}+H_0^{\rm (Au)}
-h_i^{\rm (ITO)}-h_k^{\rm (Au)},T).
\label{eq7}
\end{equation}
\noindent
Although this method is an approximate one, at short separations it
leads to the same results for the roughness corrections to the Casimir
force arising due to stochastic roughness as the more fundamental
scattering approach.\cite{36a}
In our experiment, the roughness correction reaches the maximum
value of 2.2\% at $a=60\,$nm and becomes less than 1\% and 0.5\%
at separations $a\geq 90\,$nm and $a\geq 116\,$nm, respectively.
When separation increases to about the correlation length of surface
roughness (which is equal
to a few hundred nanometers), the method of geometrical
averaging underestimates the roughness correction. In this separation
region the scattering approach is more exact. Note, however, that at
such large separations the effect of roughness in negligibly small
and can be disregarded.
Thus, for practical purposes the method of geometrical averaging
is quite sufficient when dealing with stochastic roughness
(for periodically corrugated boundary surfaces with sufficiently
small period it is necessary to use the scattering approach to obtain
good agreement between experiment and theory.\cite{36b,36c}

The computational results for $F^{\rm theor}$ as a function of $a$ are
shown by the pairs of solid lines in the main fields of Fig.~3(a) for
the untreated sample and Fig.~3(b) for the UV-treated sample.
Two different lines correspond to different extrapolations of
${\rm Im}\,\varepsilon^{\rm (ITO)}$ to high frequencies indicated in
Fig.~4(a,b) by the short-dashed lines. We emphasize that the two
solid lines in Fig.~3(a) were computed with the dielectric permittivity
shown by the two solid lines in Fig.~5(a), i.e., with included
contribution of free charge carriers of the untreated sample.
The theoretical lines in Fig.~3(a) form the theoretical band which is
in very good agreement with the experimental data indicated as crosses.
This result is rather natural and confirms that the optical properties
of the ITO film were measured correctly.

A different situation arises with Fig.~3(b) where the theoretical
forces  were computed with the dielectric permittivity
shown by the two solid lines in Fig.~5(b), i.e., with the
contribution of free charge carriers of the UV-treated sample
disregarded.
The theoretical band formed by the two solid lines in Fig.~3(b) is
also in very good agreement with the experimental data indicated
as crosses, but the reason for this agreement may seem unclear.
The point is that the optical and electric properties of the
untreated and UV-treated ITO samples at laboratory temperature are
shown to be very close. This suggests the use of the two dashed
lines in Fig.~5(b) as the true dielectric permittivity of the
UV-treated ITO sample. The computational results obtained in this way
are presented by the two dashed lines in Fig.~7(a) where the same
experimental data as in Fig.~3(b) are indicated as crosses.
As is seen in Fig.~7(a), the obvious natural computational
approach results in complete
contradiction with the data, whereas the seemingly
ad hoc omission of the contribution from free charge carriers brings
theory into very good agreement with the data. Note that almost the same
computational results are obtained when the free charge carriers are
described by the plasma model with the plasma frequency $\omega_p=9.0\,$eV
for Au and with the so-called longitudinal\cite{37} plasma frequency
$\omega_p=1.3\,$eV for ITO (which is smaller than $\omega_p$ used
for extrapolation of the measured optical data to lower frequencies).
\begin{figure*}[t]
\vspace*{-7.7cm}
\centerline{\hspace*{3.2cm}\psfig{file=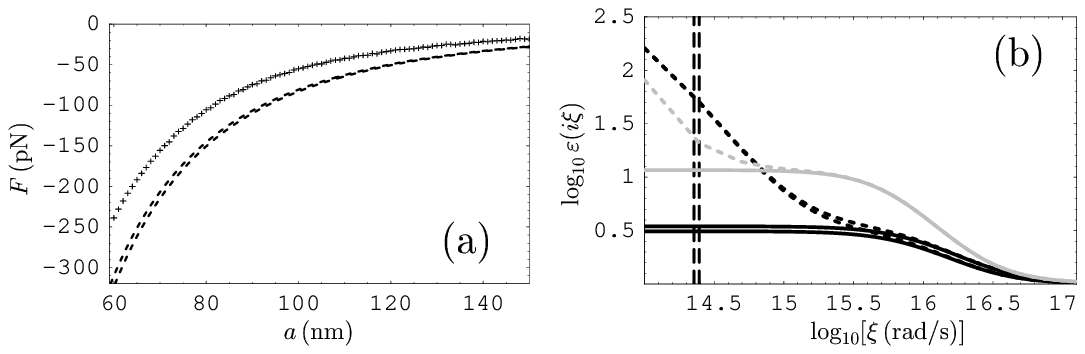,width=24cm}}
\vspace*{-22.5cm}
\caption{(a) The mean measured Casimir force as a function of
separation (crosses) and the theoretical results computed with
included contribution of free charge carriers (two dashed lines)
for the UV-treated sample.
(b) The pairs of solid black lines and short-dashed black lines
show the dielectric permittivity of the UV-treated ITO as a function
of imaginary frequency with omitted and included contribution
of free charge carriers, respectively.
The solid grey and short-dashed grey lines
show the dielectric permittivity of Si in the presence of laser light
 with omitted and included contribution
of free charge carriers, respectively.
The vertical long-dashed lines indicate the first Matsubara frequency
at $2^{\circ}$C (left) and $27^{\circ}$C (right).}
\end{figure*}

One way of looking at the decrease in the magnitude of the Casimir
force under the UV treatment is in terms of a phase transition.
It can be assumed that the UV treatment of the ITO results in
the Mott-Anderson
transition from a metallic to dielectric state without noticeable
change of the optical and electrical properties at room temperature.
Specifically, the resistivity of both the untreated and UV-treated
samples was measured and shown to be unchanged.
The assumption of a phase transition
finds  support in the observation that the UV
treatment of ITO leads to a lower mobility of charge carriers.\cite{38}
The dielectric nature of the UV-treated ITO can be verified by the
investigation of its electrical properties at low temperature.
If the phase transition to dielectric state has occurred,
the conductivity of the ITO sample should vanish with temperature.
Then the present experiment becomes comparable with other
experiments in the Casimir physics which also confirm that the
contribution of free charge carriers of dielectric materials
should be disregarded. We discuss this point in more detail
in the next section.

\section{Phase Transition from Metallic to Dielectric State and
Reduction in the Casimir Force}

In the experiment\cite{14,15} the difference in the Casimir force
between an Au-coated sphere and Si plate in the presence and in the
absence of laser pulses on the plate has been measured
at room temperature. In the absence of a laser pulse Si was in
dielectric state with the density
of charge carriers $n\approx 5\times 10^{14}\,\mbox{cm}^{-3}$.
In the presence of a laser pulse Si was in metallic state because
the density of charge carriers was up to 5 orders of magnitude
higher. Thus, in the presence of laser pulse charge carrier density
was higher than the critical value $n_{cr}$ and in the absence
of laser pulse Si had $n\ll n_{cr}$.
The theoretical results for the difference Casimir force computed using
the Lifshitz theory were found to be in agreement with experimental
data only when the free charge carriers in dielectric Si
(i.e., in the absence of laser pulse) were disregarded.
This is similar to the present experiment. The measured
difference in the Casimir forces before and after the phase
transition was determined by the actual dielectric permittivity
of metallic Si. However, to achieve agreement with the data, the
dielectric Si was replaced with some perfect isolator of zero
conductivity, as done with the UV-treated ITO.

In spite of some similarity, there are also differences between
the two experiments. In the experiment with ITO the dielectric
permittivities of the untreated and UV-treated samples are almost
the same. As to the Si plate, the contributions of charge carriers
in the absence and in the presence of laser light are quite
different because charge carrier densities differ by almost 5 orders
of magnitude. One more dissimilarity is that in the experiment
with an illuminated Si only a few percent increase in the magnitude
of the Casimir force was achieved in the presence of  laser pulse
(for instance, a 3.8\% and 5.6\% increase at $a=100$ and 150\,nm,
respectively\cite{4}). This is much smaller than the reduction in
the force magnitude observed in the ITO sample after the UV treatment.

As to the first difference between the two experiments, our measurement
with the UV-treated ITO is really
the first demonstration of the significant
decrease in the magnitude of the Casimir force with no
apparent change of
dielectric permittivity. However, if the experiment\cite{14,15} with
an illuminated Si would be repeated with a sample having the density of
charge carriers $n$ in the dark phase only
slightly smaller than $n_{cr}$,
almost
the same measurement results, as in Refs.~\refcite{14} and \refcite{15},
are expected. In this case a few percent difference in the Casimir
forces would be achieved with almost no change in the dielectric
permittivity. A similar effect of the change in Casimir force with
almost no change of dielectric permittivity can be also observed in
the proposed experiment with  a patterned plate\cite{39} consisting
of two halves made of semiconductor with $n$ only slightly smaller
than $n_{cr}$ and
$n$ only slightly larger than $n_{cr}$ (i.e., in the dielectric and
metallic states, respectively).

The second difference between the two experiments mentioned above
is simply explained by the different materials chosen. In Fig.~7(b)
we plot the dielectric permittivity of the UV-treated ITO with
omitted (the two solid black lines) and included (the two short-dashed
black lines) free charge carriers as a function of imaginary
frequency on a double logarithmic scale.
In the same figure the solid grey  and short-dashed grey lines show the
permittivity of dielectric Si with omitted contribution of charge
carriers and metallic Si in the presence of laser light,
respectively.
As is seen in Fig.~7(b), the inclusion of free charge carriers leads
to relative increase of the dielectric permittivity calculated at the
respective first Matsubara frequencies by a factor of about 17 for ITO
and by a much smaller factor of about 1.8 for Si.
This explains why using the UV-treated ITO allows one to obtain so large
a reduction in the magnitude of the Casimir force as reported here.

\section{Conclusions and Discussion}

In this paper we presented additional experimental and theoretical
information on the recently observed effect of reduction in the
Casimir force without change of dielectric permittivity.\cite{18}
The reduction occurs after the ITO plate interacting with Au sphere
undergoes  UV treatment. In so doing the dielectric permittivity
of ITO measured by means of ellipsometry both before and after the
UV treatment remains almost the same. We have shown that this effect
can be described in the framework of the Lifshitz theory if the
contribution of free charge carriers in the dielectric permittivity
of the UV-treated sample is omitted. Although the presented facts may
seem paradoxical, they find a natural explanation under the
assumption that the UV treatment caused the transition of an ITO
sample from metallic to dielectric phase.
Such transition can occur without significant changes of the
dielectric permittivity at room temperature (for instance, dielectric
to metal transition in doped semiconductors with increase of doping
concentration above the critical value). Then  neglecting the
contribution of free charge carriers in the UV-treated ITO becomes
similar to the neglect of dc conductivity in dielectric materials
in the Lifshitz theory. As discussed above, the inclusion of dc
conductivity of dielectric Si in the absence of laser pulse was found
to be in contradiction with measurements of the Casimir force between
an Au sphere and Si plate.\cite{14,15} In one more experiment
it was demonstrated\cite{40} that the measured Casimir-Polder force
between ${}^{87}$Rb atoms and dielectric (SiO${}_2$) plate is consistent with
theoretical predictions with the dc conductivity of SiO${}_2$ omitted,
but disagree with the theoretical prediction taking dc conductivity into
account.\cite{41}

Thus, several experiments performed with dielectric
test bodies including the one considered in this paper support the
phenomenological prescription\cite{1,2,46} on how to apply
the Lifshitz theory to dielectric materials with no contradictions
with thermodynamics and experiment (the inclusion of dc conductivity of
dielectrics in the Lifshitz theory was shown\cite{47} to violate
the Nernst heat theorem). According to this prescription,
in calculations of dispersion forces between
 dielectrics and semiconductors of dielectric type free charge
carriers should be disregarded.
The possibility to obtain significantly different Casimir forces from
the test bodies with nearly equal dielectric permittivities,
as observed in the present experiment, is an immediate consequence
of this prescription (see the discussion of the proposed
experiment\cite{39} in Sec.~4).
It is obvious, however, that such phenomenological prescriptions
require a more fundamental theoretical justification.

Regardless of whether the above theoretical explanation will be
confirmed in future, the observed effect of reduction in the
Casimir force without change of dielectric permittivity is
prospective for applications in nanotechnology, specifically, in
connection with problems of lubrication and stiction.
The Casimir and van der Waals forces lead to collapse of closely spaced
moving parts of nanoelectromechanical systems to the fixed
electrodes leading to loss of functionality of devices.
Significant reduction in the magnitude of the Casimir force may help
to increase the output of efficient nanosystems.

\section*{Acknowledgments}

This work was supported by the DARPA Grant under Contract
No.~S-000354 (equipment, A.B., R.C.-G., U.M.),  NSF Grant
No.~PHY0970161 (C.-C.C., G.L.K., V.M.M., U.M.) and DOE Grant
No.~DEF010204ER46131 (G.L.K., V.M.M., U.M.).
G.L.K.\ and V.M.M.\ were also partially supported by the DFG Grant
BO\ 1112/20-1. U.M., G.L.K.\ and V.M.M.\ are grateful to the local
Orginizing Committee of QFEXT11 (Benasque, Spain) for their kind
hospitality.

\end{document}